     \documentstyle[proceedings,psfig]{./crckapb} 
\begin{opening}

   \title{Three-dimensional Radiative Transfer\protect\\
    with Multilevel Atoms}

   \author{P. Fabiani Bendicho}

   \author{J. Trujillo Bueno}

   \institute{Instituto de Astrof\'\i sica de Canarias,\protect\\
    E-38200, La Laguna 
   Tenerife, Spain}

\end{opening}

\runningtitle{Three-dimensional Radiative Transfer}

\begin{document}

%%%\begin{document}

% The \begin{document} command comes after the \end{opening}
% command.

\begin{abstract}\footnotesize\ 
The efficient numerical solution of Non-LTE multilevel transfer problems requires the combination of highly convergent iterative schemes with fast and accurate formal solution methods of the radiative transfer
(RT) equation. This contribution\footnote{Published in 1999 in the book {\it Solar Polarization}, edited by K.N. Nagendra \& J.O. Stenflo. Kluwer Academic Publishers, 1999. (Astrophysics and Space Science Library ; Vol. 243), p. 219-230} begins presenting
a method for the formal solution
of the RT equation in three-dimensional (3D) media
with horizontal periodic boundary conditions.
This formal solver is suitable
for both, unpolarized and polarized 3D radiative transfer 
and it can be easily combined with the iterative schemes 
for solving non-LTE multilevel transfer problems
that we have developed over the last few years. 
We demonstrate this by showing some schematic 3D multilevel
calculations that illustrate the physical effects of horizontal
radiative transfer. These Non-LTE calculations have been carried out
with our code MUGA 3D, a 3D multilevel Non-LTE code based on
the Gauss-Seidel iterative scheme that Trujillo Bueno
and Fabiani Bendicho (1995) developed for RT applications.


\end{abstract}

\section{Introduction}

To what extent can we trust diagnostic results obtained with the assumption
that the solar atmospheric plasma is composed of {\it homogeneous}
plane-parallel layers or via approximations that neglect {\it horizontal}
radiative transfer (RT) effects? How important are the errors
in the magnetic fields, temperatures and velocities inferred by
confronting spectro-polarimetric observations with Non-LTE 
1D RT model calculations? Clearly, to provide proper answers
to questions like these requires 
to develop first efficient 3D
RT methods that allow Non-LTE effects in complex atomic models with many levels
and transitions to be rigorously investigated. 

There is a second reason which makes the development of fast
iterative methods for 3D Non-LTE RT so relevant. This is because
processes of energy exchange by radiation play an important role
in the structure and dynamical behaviour of the stellar magnetized plasma.
Thus, for instance, if one wishes to perform time-dependent 
radiation hydrodynamics
simulations similar to those carried out by Carlsson and Stein (1997),
but in 3D instead of 1D, it turns out to be imperative to have first access
to numerical methods capable of accurately yielding the self-consistent
atomic level populations at the cost of only {\it very} few formal 
solution times. 

The efficient solution of multilevel transfer problems
requires the combination of a highly convergent iterative scheme with a fast
formal solver of the RT equation.
In Section 2 we briefly comment on a hierarchy of iterative schemes
that can be applied for solving multilevel Non-LTE problems with increasing
improvements in the convergence rate and total computational work. The
3D multilevel transfer calculations that we present 
in this contribution have been obtained by combining
a highly convergent iterative scheme based on Gauss-Seidel iteration
(Trujillo Bueno and Fabiani Bendicho, 1995) with a fast 3D formal
solver that has parabolic accuracy
(see Section 3). As was the case
with our 2D formal solver, our generalization to 3D is
based on the ``short-characteristics'' method of Kunasz and Auer (1988).
Our 3D multilevel code
is called MUGA 3D (``Multi-level Gauss-Seidel Method'') and it
is substantially {\it faster} than our code MALI 3D, which is based
on Jacobi iteration (see Rybicki and Hummer, 1991;
Auer, Fabiani Bendicho and Trujillo Bueno, 1994).

Section 3 briefly describes our 3D formal solver as applied to 
the scalar transfer equation for the specific
intensity (I). In order to be able
to consider 3D atmospheric models where solar plasma structures
repeat themselves along the horizontal directions
we choose horizontal periodic boundary conditions
along the Cartesian coordinates X and Y. Although we do not give
any details here, we have also generalized to 3D the Stokes-vector
1D formal solver method developed by Trujillo Bueno (1998), which
is based on the matrix exponential approximation to the evolution operator.  

In Section 4 we show some illustrative 3D multilevel
transfer calculations for a 5-level Ca II model atom where the H, K
and infrared triplet lines are treated simultaneously, taking fully
into account the {\it interlockings} by which photons are converted 
back and forth between the different line transitions
in the assumed 3D medium. Here we consider
schematic 3D solar models characterized by horizontal sinusoidal temperature
inhomogeneities. With the help of these 3D multilevel
calculations we are able to illustrate
some subtle effects of horizontal radiative transfer that are important
for the correct interpretation of high spatial resolution observations.
Finally, Section 5 gives our conclusions.
 
\section{Iterative Methods for Multilevel Transfer}

The simplest procedure one might think of to solve 
self-consistently the kinetic and RT equations is $\Lambda-$iteration: using the
current estimate of the atomic level populations at each spatial grid-point
(or, more in general, of the irreducible tensor components
of the atomic density matrix; see Trujillo Bueno 1999) 
evaluate the absorption and emission coefficients. Next, solve the radiative transfer equation and compute the radiation field intensity
in all transitions. Finally, with the radiative rates obtained, solve the kinetic equations at each spatial grid-point {\it independently} and obtain a 
new estimate of the atomic level populations. However, as is well known,
under typical NLTE conditions in optically thick media this 
$\Lambda-$ iteration method
converges extremely slowly. This is certainly regrettable because with this
method there is no need of inverting large matrices and the computing time per 
iteration is minimal. In any case, as demonstrated by Trujillo Bueno and Manso Sainz (1999), in solar-like atmospheres 
the $\Lambda-$ iteration method can be used to find the self-consistent solution of Non-LTE polarization transfer problems {\it if}
one initializes using the ``exact'' solution corresponding to the
unpolarized transfer case.

The dream of numerical RT is to develop iterative
methods where everything goes as simply as with the $\Lambda-$iteration
scheme, but for which the convergence rate is extremely high. 
The iterative methods for RT applications based on Gauss-Seidel iteration
that we have developed over the last few years
have been worked out with this motivation in mind. Their convergence
rate is {\it extremely} high, there is no need
of constructing and inverting any large matrix,
and the computing time per iteration
is similar to that of the $\Lambda-$iteration method. 
A full account of these developments 
can be found in the following publications:

1) Auer, Fabiani Bendicho and Trujillo Bueno (1994)
present the generalization of
the Jacobi-based {\bf MALI} method of Rybicki \& Hummer (1991)
to multilevel RT in 2D. They also developed
a short-characteristics strategy to do the formal solution
in 2D Cartesian coordinates with horizontal periodic boundary conditions.
Of particular interest is a simple {\it grid-doubling} technique
which both rapidly finds the converged solution in fine
meshes and automatically estimates its corresponding {\it true} error.
The total computational work scales as ${\rm NP}^2$, with NP the {\it total}
number of spatial grid-points in a fixed computational domain.

2) Trujillo Bueno and Fabiani Bendicho (1995) developed a novel
iterative scheme based on 
Gauss-Seidel (GS) iteration ({\bf MUGA}). This is the paper on which our present work is based on. The total computational work
scales as ${\rm NP}^2/4$ for the pure GS method
(implemented as suggested in the conclusions of their paper), 
and as ${\rm NP}{\sqrt{\rm NP}}$ for the 
successive overrelaxation (SOR) method. 
This paper was fundamental for a successful
development of our {\it non-linear} multigrid method for RT applications.

3) Fabiani Bendicho, Trujillo Bueno and Auer (1997) consider
the application of the {\it non-linear} multigrid method (see Hackbush, 1985) to multilevel RT. Here the iterative scheme is composed of two parts:
a {\it smoothing} one where a small number of MUGA iterations on the desired
{\it finest} grid are used to get rid of the high-frequency
spatial components of the error in the current estimate,
and a correction obtained from the solution of an 
error equation in a {\it coarser} grid. 
With this method the total computational work scales simply as NP,
although it must be said that
the computing time per iteration is about 4 times larger than that
required by the $\Lambda-$iteration method.

\begin{figure}[h]
\centerline{\psfig{file=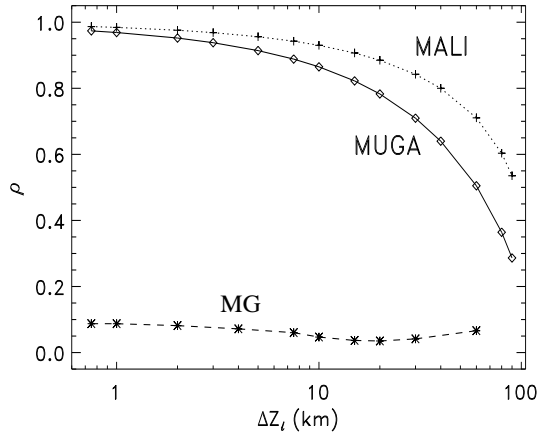,width=8cm}}
%\vspace{-4truemm}
\caption{Variation with the grid-spacing $\Delta{z}$
of the maximum eigenvalue of the iteration
operator corresponding to several multilevel iterative schemes.
The MG symbol refers to our {\it non-linear} multigrid code.}
\end{figure}

In order to compare the convergence rates of these three 
iterative methods, we
present in Fig. 1 an estimate of the maximum eigenvalues ($\rho$) of the
corresponding iteration operator,
which controls the convergence properties
of such iterative schemes (see Trujillo Bueno and Fabiani Bendicho, 1995).
The knowledge of this maximum eigenvalue ($\rho$) is useful because
errors decrease as ${\rho}^{itr}$, 
where ``$itr$'' is the iterative step. We obtain this information 
from multilevel Ca II
calculations in several 1D grids with decreasing grid-size
($\Delta{z}$) by calculating $R_c(itr+1)/R_c(itr)$
for $itr>>1$, where $R_c(itr)$ is the maximum
relative change in the level populations. As it can be noted in Fig. 1 the
convergence rate of both, the MALI and MUGA schemes decreases when the spatial
resolution of the grid is improved, while the maximum eigenvalue
of our {\it non-linear} multigrid method is always 
very small ($\rho\sim0.1$) and
{\it insensitive} to the grid-size. A maximum eigenvalue $\rho=0.1$ means that
the error decreases by one order of magnitude each time we perform an iteration.
This explains that, typically, two multigrid iterations
are sufficient to reach the self-consistent solution for the atomic level populations.

The 3D multilevel calculations shown in this contribution were
obtained with our code MUGA 3D, i.e. with a multilevel GS scheme based
on the paper by Trujillo Bueno and Fabiani Bendicho (1995) combined
with the following 3D formal solver. In practical applications we 
always use MUGA 3D with Ng (1974) acceleration.

\section{The 3D formal solver.}

The scalar RT equation for the specific intensity is

\begin{equation}
{{{\rm d}{\rm I}_{\nu}}\over{{\rm d}{s}}}\,=\,
{\chi_\nu}\,(\,{\rm S}_{\nu}\,-\,{\rm I}_{\nu}\,)\,, 
\end{equation}
where $s$ is the geometric distance along the ray propagating
in a certain direction
in a 3D medium, $\chi_\nu$ is the total opacity
and ${\rm S}_{\nu}$ the source function. 

\begin{figure}
\centerline{\psfig{file=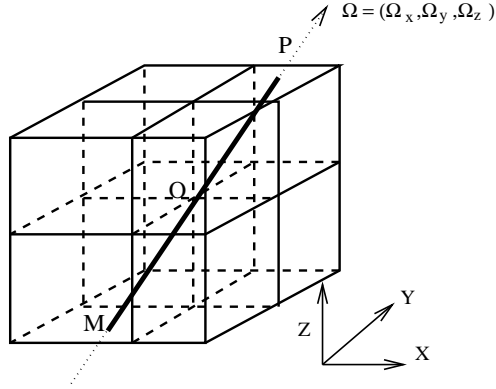,width=6.5cm}}
%\vspace{-7truemm}
\caption{3D Cartesian spatial grid surrounding the grid-point
of interest (O) where the specific intensity ${\rm I}_{\rm O}$
is to be calculated.
This is done by solving analytically the integral
of Eq. (2) along the {\it short-characteristics} MO
corresponding to the ray of direction $\vec{\Omega}$.} 
\end{figure}

We now consider a 3D 
Cartesian spatial grid (see Fig.~2). Point O is the grid-point 
of interest at which one wishes to calculate
the specific intensity ${\rm I}_{\rm O}$, for a given frequency ($\nu$) 
and a direction (${\vec{\Omega}}$).
Point M is the the intersection point
with the grid-plane that one finds when moving along $-\vec{\Omega}$.
At this {\it upwind} point we assume that the specific intensity
${\rm I}_{\rm M}$ for the same frequency and angle is known
from previous steps. In a similar way, point
P is the intersection point with the grid-plane
that one encounters when moving along 
$\vec{\Omega}$. We also introduce the optical depths
along the ray between points M and O ($\Delta{\tau}_{\rm M}$)
and between points O and P ($\Delta{\tau}_{\rm P}$). 
From the formal solution of the previous
transfer equation one finds that

\begin{equation}
{\rm I}_{\rm O}={\rm I}_{\rm M}\,{\rm e}^{-\,\Delta{\tau}_{\rm M}}\,+\,
\int_{0}^{\Delta{\tau}_{\rm M}}\,{\rm S}(t)
{\rm e}^{-({\Delta{\tau}_{\rm M}}\,-\,t)}\,dt,
\end{equation}
where the optical depth variable $t$ is measured
from M to O.

The integral of Eq. (2) can be solved {\it analytically} by
integrating along the {\it short-characteristics} MO assuming
some prescribed variations for the source function
(e.g. linear variation along M and O, or parabolic along M,O and P,
or cubic-centered around point O, etc.). Our 3D formal solution
method assumes that the source function ${\rm S}(t)$ varies {\it parabolically}
along M,O and P. The result reads:

\begin{equation}
{\rm I}_{\rm O}={\rm I}_{\rm M}\,{\rm e}^{-\Delta{\tau}_{\rm M}}\,+\,
{\Psi}_{\rm M}{\rm S}_{\rm M}\,+\,{\Psi}_{\rm O}{\rm S}_{\rm O}\,+\,
{\Psi}_{\rm P}{\rm S}_{\rm P},
\end{equation}
where $\Psi_{\rm X}$ (with X either M, O or P)
are given in terms of the quantities $\Delta{\tau}_{\rm M}$ and
$\Delta{\tau}_{\rm P}$ that we evaluate numerically by assuming
that ${\rm ln}(\chi)$ varies linearly with the geometrical depth, $\chi$ 
being the opacity.

It is very important to point out that one should avoid
the use of a formal solution method based on a 
linear interpolation formula, i.e. one should avoid assuming that, for each
grid-point O of interest, the source function
varies linearly along points M and O. 
Otherwise, the accuracy of the self-consistent
solution will never be better than about 10$\%$,
even by choosing a very large number of grid-points per 
opacity scale height (see Trujillo Bueno, 1998).
The reason for this is that the use of linear interpolation 
for ${\rm S}(t)$ leads to formal solution methods that are
unable to yield the correct asymptotic behaviour for the intensity
when having nonlinear source functions in optically thick atmospheres.
We emphasize that our 3D formal solution
method is based on the above-mentioned parabolic approximation and
it only uses the linear approximation formula 
for calculating the radiation field
at the upper and lower boundaries for rays going out of such boundaries. 
However, as we illustrate below, the use of the parabolic approximation for
investigating problems where we have sudden variations in the physical
quantities requires to implement it using 
an improved version of the monotonic upwind interpolation
technique applied by Auer and Paletou (1994). 

The application of this formal solution method in 1D is straightforward.
A detailed description of how to implement it in 2D slabs
with prescribed irradiation on the lateral boundaries can be
found in Kunasz and Auer (1988) and Auer and Paletou (1994).
In 2D with horizontal periodic boundary conditions is 
slightly more complicated and a suitable strategy has been described by 
Auer, Fabiani Bendicho and Trujillo Bueno (1994).

The main changes when going to 3D 
imposing horizontal periodic boundary conditions
lie in the interpolation. We have
assumed that $\rm I_M$ is known (see Fig.~2) but, in most cases,
the M-point (like the point P)
will not be a grid-point of the chosen 3D spatial grid. The
intensity at this M-point has to be calculated
by interpolating from the available information at the nine surrounding
grid-points, as we must also do for obtaining the opacities
and source functions at M and P.

Parabolic interpolation can however generate spurious
negative intensities if the spatial variation of the physical quantities
is not well resolved by the spatial grid. This happens,
for instance, if one tries to 
simulate the propagation of a beam in
vacuum using a three dimensional grid. The analytical solution in this
case is simply $\rm I_O=I_M$, and any numerical error is due to the
effect of the conventional way of applying the parabolic interpolation. 
To avoid these problems we have improved the 1D monotonic interpolation
strategy of Auer and Paletou (1994), and generalized it 
to the two-dimensional parabolic interpolation that is required for
3D RT calculations (Fabiani Bendicho and Trujillo Bueno, in preparation).

\begin{figure}[ht]
 {\psfig{figure=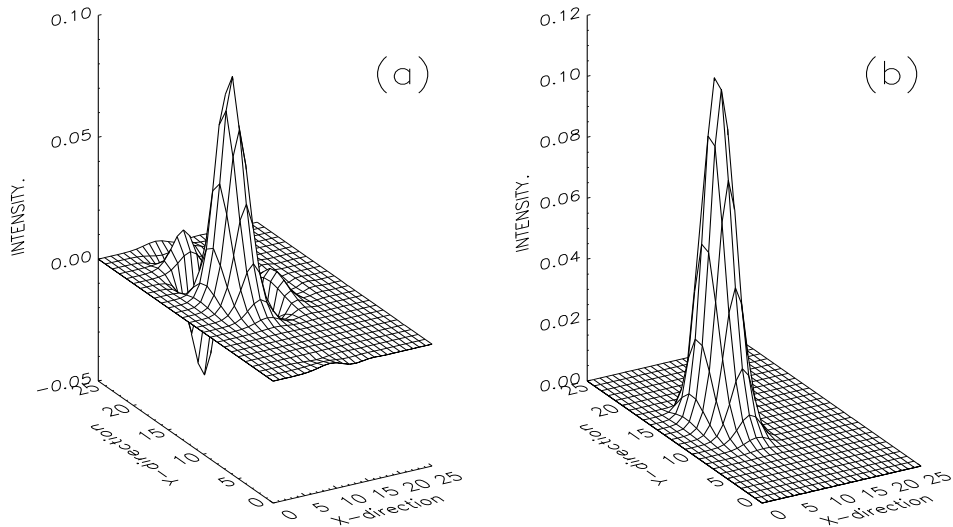,height=8cm,width=13cm}}
\vspace{-7truemm}
\caption{Ray propagating in vacuum. Emergent intensity with parabolic
interpolation implemented
without (a) and with (b) our improved version of the 
monotonic upwind interpolation strategy.}
\end{figure}

In Fig.~3 we show the intensity that emerges from a 3D computational box 
after having illuminated its lower boundary 
with a beam  of given intensity. Here the aim is to simulate
a beam propagating in vacuum.
Using standard parabolic interpolation (Fig.~3a)
leads to unphysical negative intensities. However, with 
our improved version of the upwind monotonic interpolation strategy
we guarantee that the parabolic interpolation is not introducing spurious
sources and sinks and its second order accuracy is maintained.

For solving Non-LTE radiative transfer 
problems as shown above the chosen formal solution method
has to be implemented in a way such that it rapidly computes,
at each spatial grid-point, the radiation field for each frequency
and direction of the chosen numerical quadratures
together with the diagonal element of the $\Lambda-$operator.
To this end, one may also try with higher-order methods if desired. We have
found that our parabolic formal solvers are fast and accurate enough, stable
and easy to work with independently of the geometry and of the
iterative method used.

\section{Illustrative 3D multilevel results}

Our aim here is to study a few illustrative results for Ca II
in a schematic 3D atmospheric model characterized by the following
temperature structure:

\begin{equation}
\rm T(X,Y,Z)\,=\,5000\,+\,500\,sin(a\,X)\,\,sin(b\,Y),
\end{equation}
where $\rm a=2\pi/P_{x}$ and $\rm b=2\pi/P_{y}$, with ${\rm P}_{x}$
and $\rm P_{y}$ the horizontal periods along the X and Y Cartesian coordinates,
respectively. For this example we choose
$\rm P_x\,=\,P_y\,=1000$ km. We also assume that the total hydrogen number 
density is exponentially stratified 
along the vertical direction (Z) with a scale height $\cal H$=100 km.
We adopt a standard 5-level Ca II atomic model and a constant electron density
$n_e=10^{11}\,{\rm cm}^{-3}$. We compare our full 3D multilevel results
with the corresponding 2D multilevel calculation and with 
results obtained using 
a well-known approximation that neglects the effect of horizontal
RT on the level populations, i.e. with the so-called
$1.5{\rm D}$ approximation (Mihalas, Auer and Mihalas, 1978; 
Solanki, Steiner and Uitenbroek, 1991).

We point out that our calculations do take into account the change 
in the line opacities caused by the assumed 500 K 
horizontal temperature inhomogeneities. Thus, the
3D multilevel transfer effects are the result of the combined action
of the {\it smoothing} effect of the source-function fluctuations
(Spiegel, 1957; Kneer and Trujillo Bueno, 1987) and the {\it channelling}
effect due to the opacity fluctuations (Cannon, 1970; Trujillo Bueno and Kneer, 
1990). Since we use a 5-level atomic model, we can examine the 3D effects
for the H, K and the infrared triplet lines. 
Figures 4, 5 and 6 give the variation with X and Y of
the vertically emergent intensity at the continuum frequency (upper part)
and at the line core (lower part). We only show this for
the H and 8662 ${\rm \AA}$ lines since for the K line and the other two infrared
lines very similar results were found.

\begin{figure}[ht]
\begin {tabular}{cc}
\hspace{-15truemm}
 {\psfig{figure=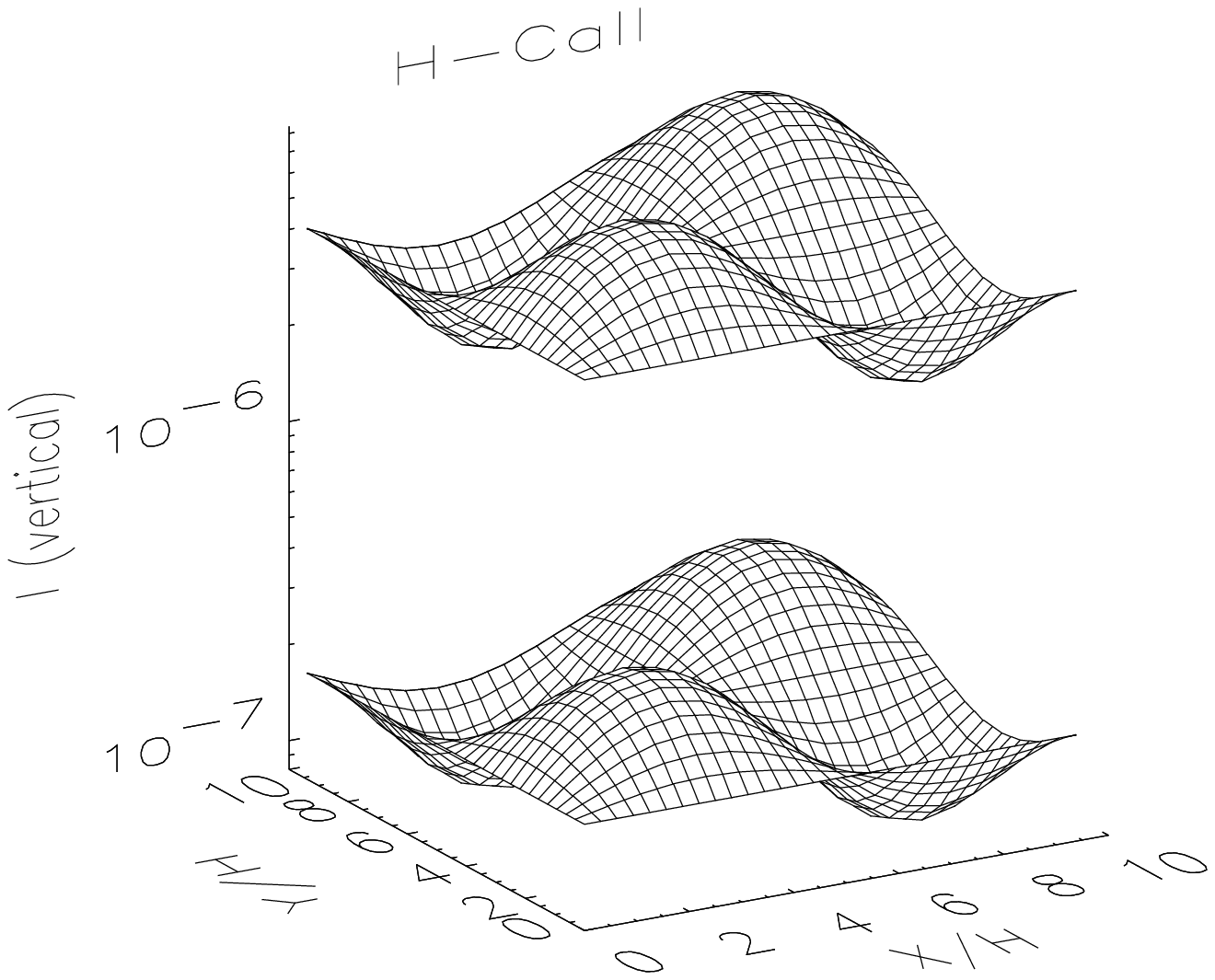,width=8.5cm}}
&
\hspace{-25truemm} 
 {\psfig{figure=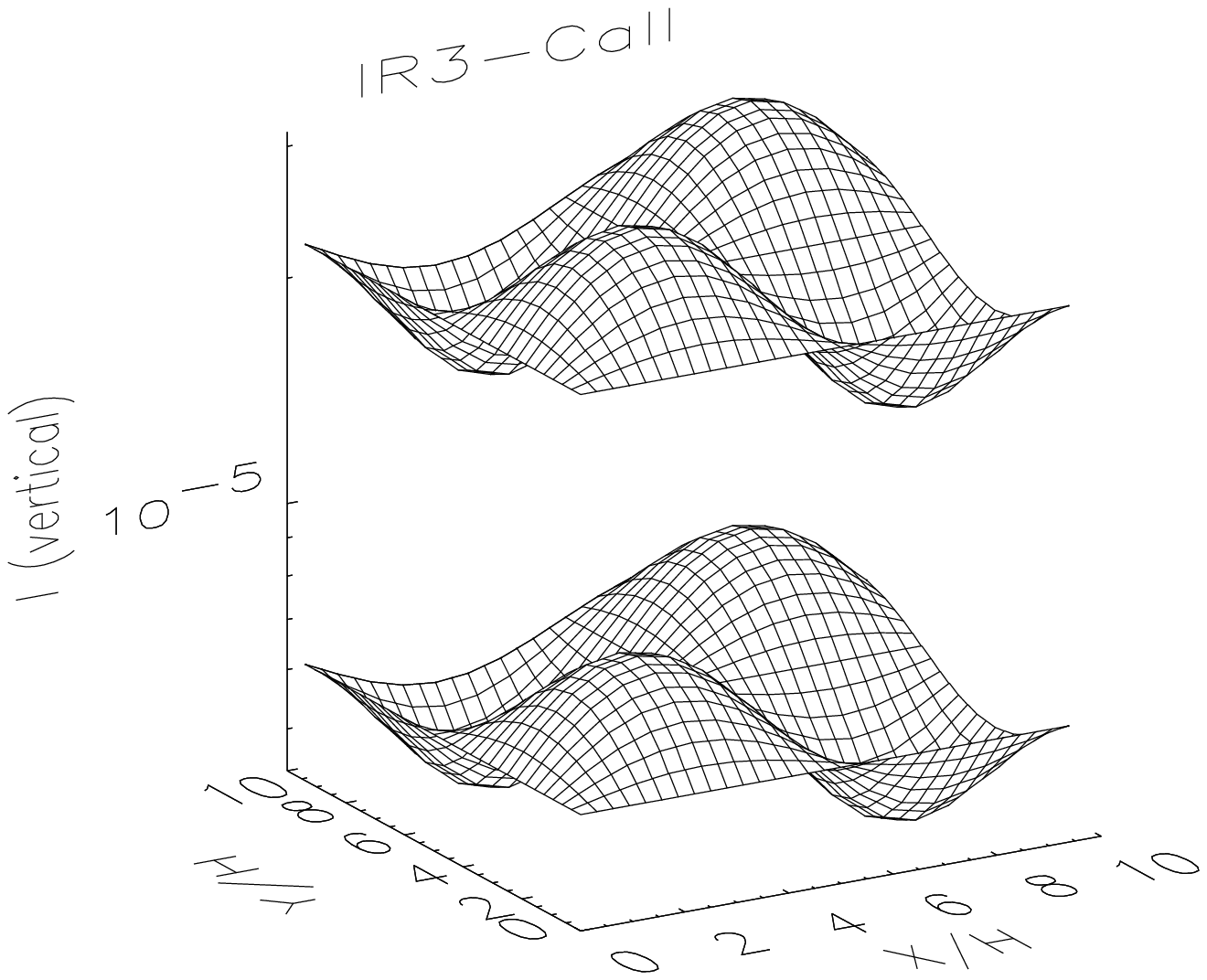,width=8.5cm}}
\end{tabular}
\vspace{-7truemm}
\caption{The vertically emergent intensity at the line core (lower 
surface plot)
and at the continuum frequency (higher 
surface plot) corresponding to the $1.5{\rm D}$
approximation applied to a model atmosphere
with ${\rm P_x=P_y}=1000$ km. These intensities
are shown for the H and 8662~${\rm \AA}$ Ca II lines in the left and
right panels of the figure, respectively.
Geometrical distances are measured in units of the 
opacity scale height $\cal H$.}
\end{figure}

Fig.~4 corresponds to calculations carried out with the $1.5{\rm D}$
approximation. As pointed out above 
this approximation completely neglects the effects
of horizontal RT on the level populations. Consequently, 
the result is that, not only
at the continuum frequency, but also at the line center the emergent
intensity shows large horizontal fluctuations. The 
$1.5{\rm D}$ atomic level populations have 
large horizontal fluctuations
at all atmospheric depths, simply because one is here neglecting
the smoothing and channelling horizontal 3D RT effects 
when solving the kinetic and RT equations.

\begin{figure}[ht]
\begin {tabular}{c c}
\hspace{-15truemm}
 {\psfig{figure=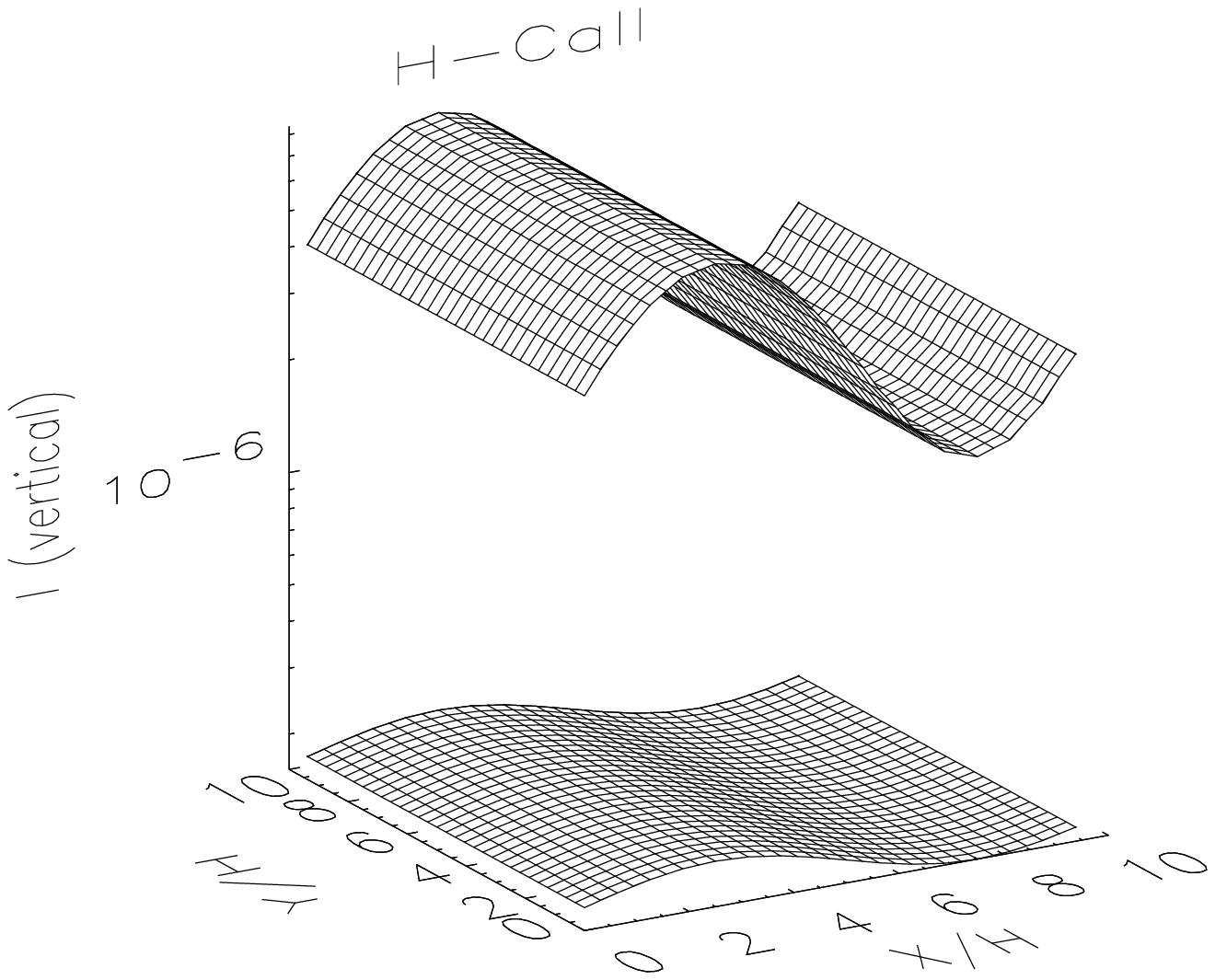,width=8.5cm}}&
\hspace{-25truemm} 
 {\psfig{figure=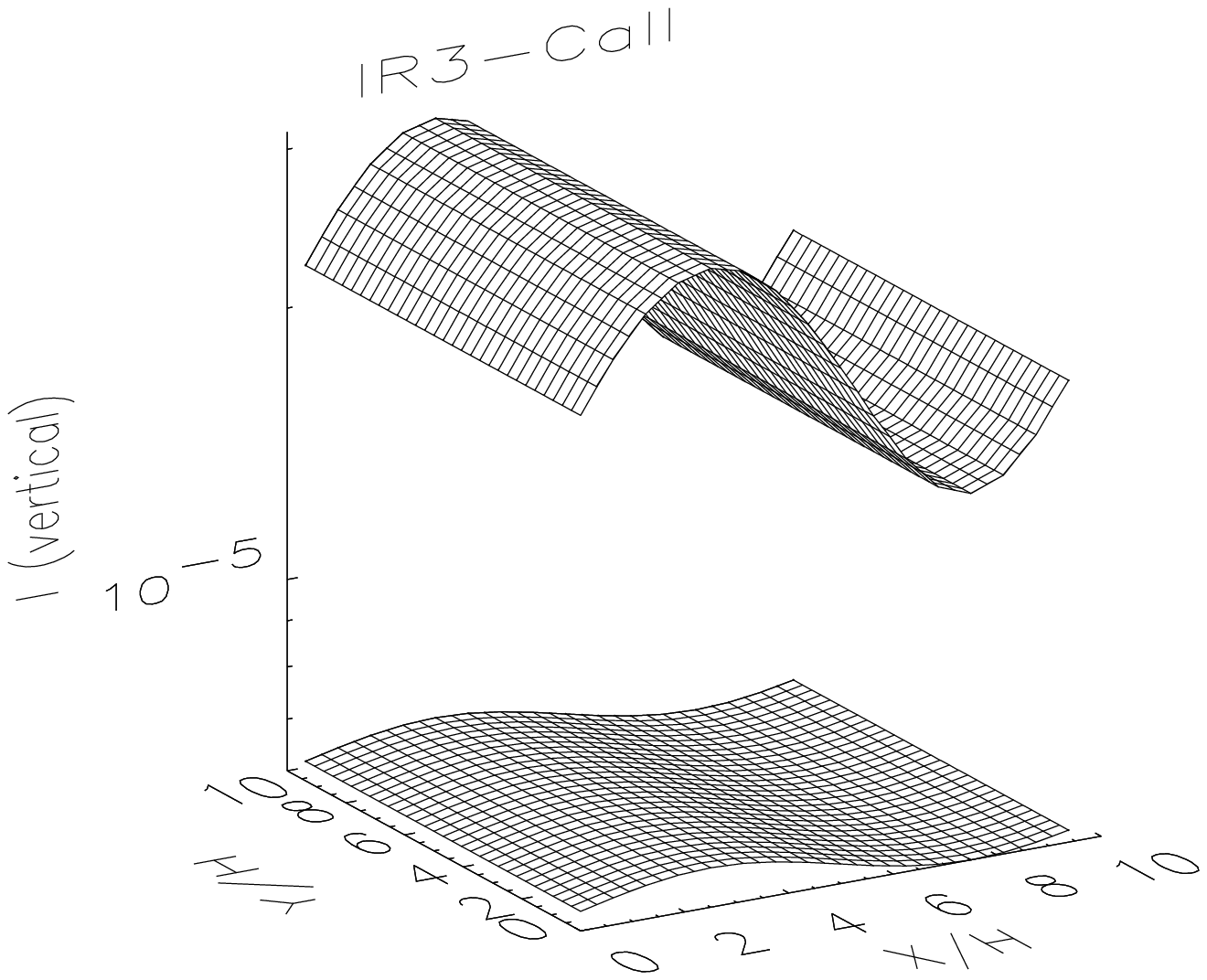,width=8.5cm}}
\end{tabular}
\vspace{-7truemm}
\caption{2D vertically emergent intensities. For more information
see Fig. 4.}
\end{figure}

Fig.~5 shows the 2D multilevel transfer results 
(see also Auer, Fabiani Bendicho
and Trujillo Bueno, 1994). Here it was assumed that the model's temperature
was only fluctuating along the horizontal X-direction
with $\rm P_x=1000$ km; i.e. horizontal transfer effects along the Y-direction  
are completely neglected. The explanation of the 
results is the following:
above the height of thermalization, 
2D horizontal transfer effects are efficiently damping the horizontal 
fluctuations of the populations of the
two uppermost levels of the chosen 5-level Ca II model atom. 
This smoothing is translated to the line source functions, which are 
proportional to 
$n_4A_{4,1}$ for the H line and to $n_4A_{4,2}$ 
for the 8662 ${\rm \AA}$ line, with $A$ the Einstein coefficient for spontaneous emission. Thus, since the continuum photons come 
from the deepest layers, one sees strong
horizontal intensity fluctuations at continuum 
frequencies, while at the line core
one finds horizontal intensity fluctuations 
of very small amplitude because these
photons come from the higher atmospheric layers. 
Note that, both for the H and the 
8662 ${\rm \AA}$ infrared lines, the vertically emergent
intensities are fluctuating in phase with the assumed thermal inhomogeneities.

\begin{figure}[ht]
\begin {tabular}{c c}
\hspace{-15truemm}
 {\psfig{figure=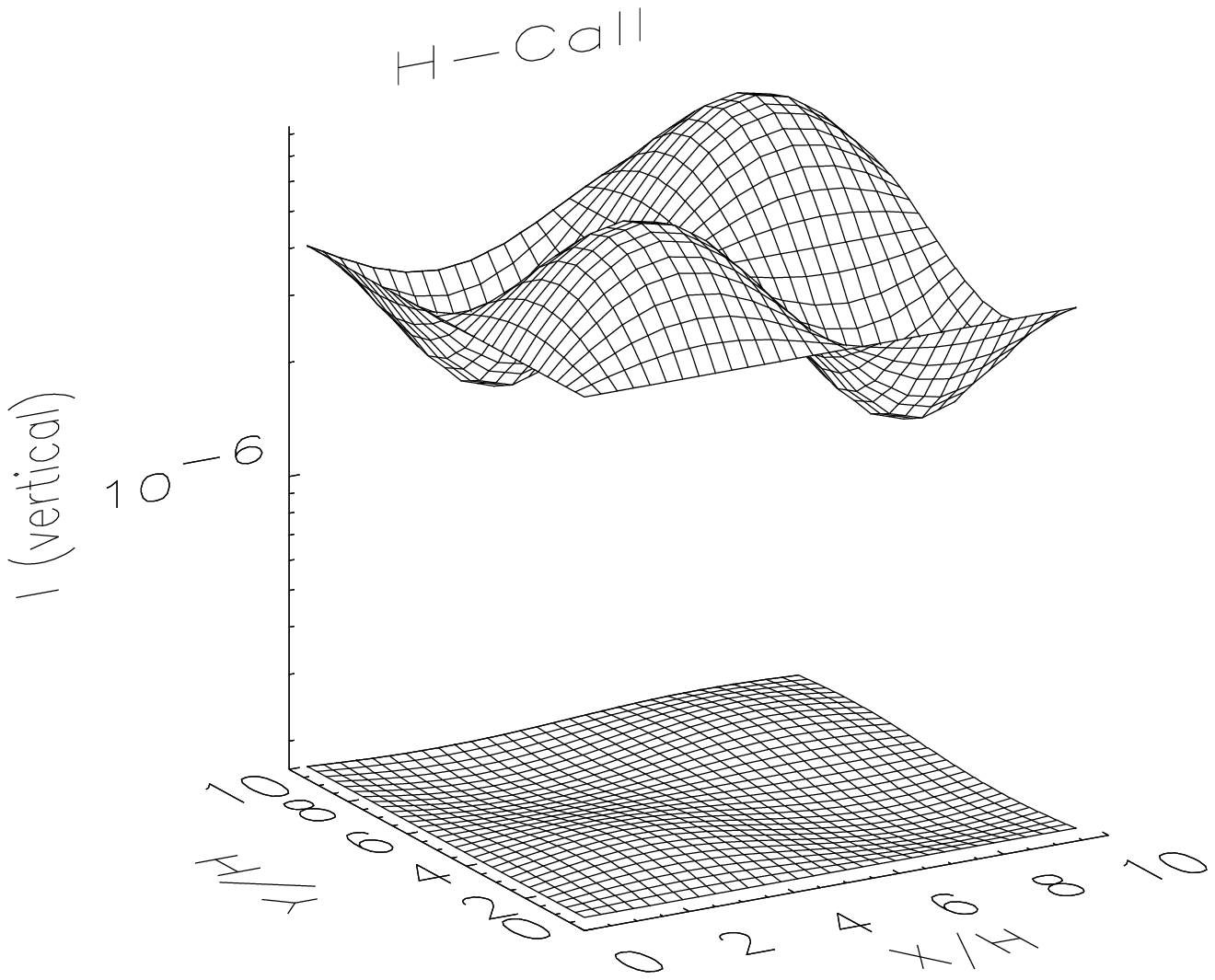,width=8.5cm}}&
\hspace{-25truemm} 
 {\psfig{figure=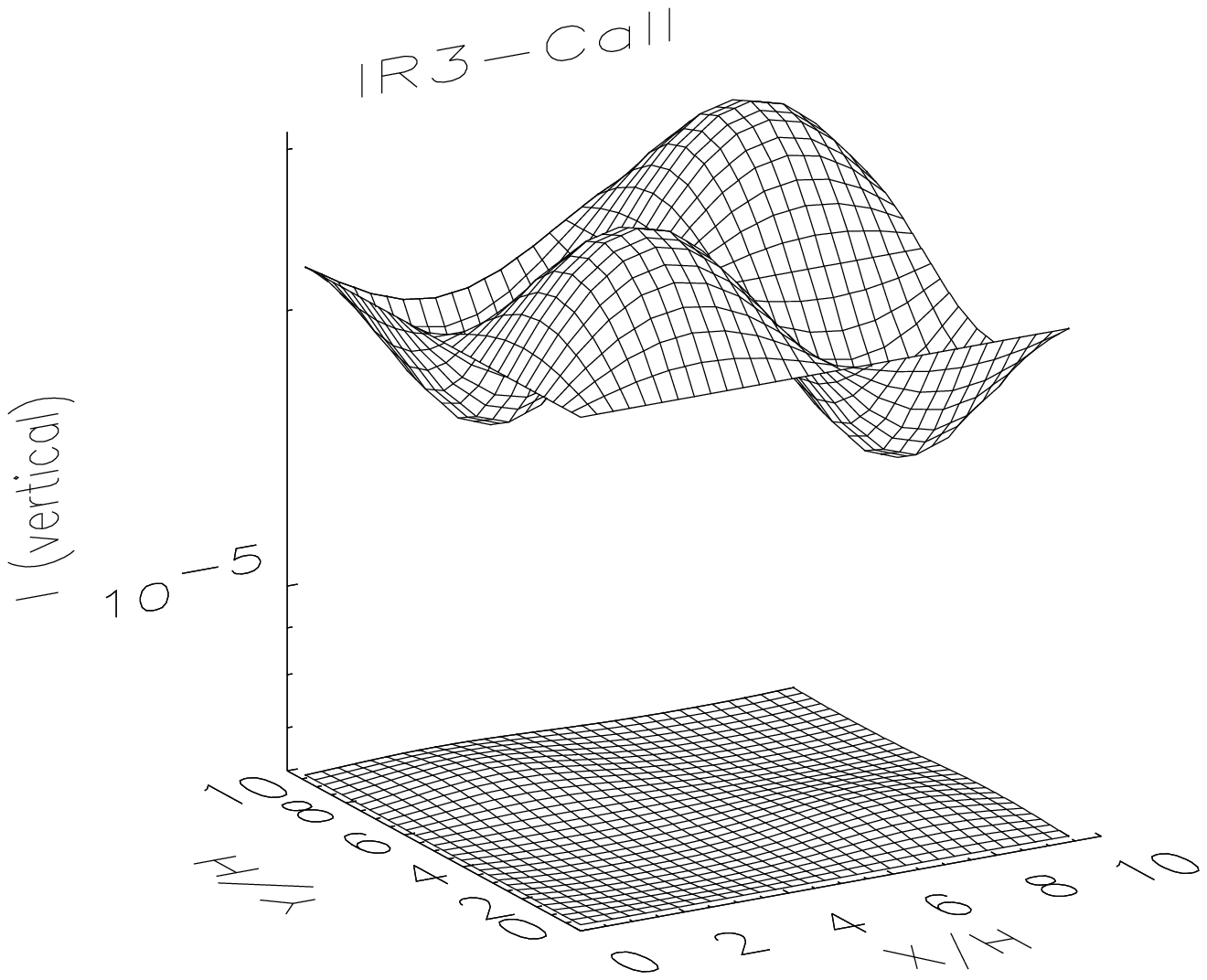,width=8.5cm}}
\end{tabular}
\vspace{-7truemm}
\caption{3D vertically emergent intensities. For more information
see Fig. 4.}
\end{figure}

Fig.~6 depicts the results of our full 3D multilevel calculation
with  $\rm P_x\,=\, P_y=1000$ km. For the H line the result is qualitatively
similar to the 2D results, although the amplitude of the horizontal 
intensity fluctuations at line-center is significantly smaller, as expected
from the fact that in 3D the horizontal transfer effects are enhanced with 
respect to those found for the 2D case. However, for the 8662 ${\rm \AA}$ line
we find a result that, at first sight, might appear surprising: at line center
the resulting horizontal fluctuation of the vertically emergent
intensities is anticorrelated with 
respect to the fluctuation of the assumed thermal inhomogeneities. 
In other words, at the line core of this infrared line,
{\it one gets more intensity from the coolest 
atmospheric points than from the hottest
ones}. The explanation is that the IR line source function shows a change,
by $\pi$, in phase above the height ($h_{c}$) in the atmosphere where the 
amplitude of the horizontal fluctuation in $n_4$ is greatly diminished.
And this change of phase is, in turn, due to
the fact that the line source function is the result of dividing
an almost non-fluctuating emissivity by a line opacity (essentially
set by $n_2$) that fluctuates in phase with the 
assumed temperature inhomogeneities. This occurs both in 2D and in 3D,
but with the difference that $h_c^{2D}\,>\,h_c^{3D}$ for given values
of $\rm P_x$ and $\rm P_y$. In fact, one finds indeed a similar result
in 2D, but for $\rm P_x$ values 
significantly smaller than 1000 km. Since the horizontal
transfer effects are more important in 3D than in 2D, it turns out that
such an effect already sets in at $\rm P_x=P_y=1000$ km.

We thus see that our 3D multilevel calculations demonstrate
the efficiency of horizontal RT effects and confirm (see Kneer, 1981;
Trujillo Bueno, 1990) that the interpretation of high-spatial resolution
observations ignoring the existence of horizontal transfer effects
(as is done when using the $1.5{\rm D}$ approximation) may substantially
underestimate the actual inhomoheneities present in the solar atmospheric plasma. 

\section{Concluding Remarks}

We have developed a 3D multilevel code (MUGA-3D) that combines
the Gauss-Seidel iterative scheme of Trujillo Bueno
and Fabiani Bendicho (1995) 
with a 3D formal solver that uses horizontal periodic boundary conditions. 
With this new code we have performed some 3D multilevel simulations that highlight the importance of carefully investigating the effects
of horizontal radiative transfer using realistic atmospheric
and atomic models. 

We point out that our 3D formal solver can be used not only for solving 
``unpolarized'' multilevel transfer problems, 
but also resonance line polarization
and Hanle effect problems, like those considered in these Proceedings
by Manso Sainz and Trujillo Bueno (1999), Paletou {\it et. al.} (1999)
or Dittmann (1999). This is because, for these
polarization transfer cases, the absorption matrix is diagonal. As a result,
we have similar equations for the Stokes I, Q and U parameters.

However, for the solution of more general polarization transfer problems,
like the Non-LTE Zeeman line transfer case
considered by Trujillo Bueno and Landi Degl'Innocenti (1996), but
in 3D instead of 1D, one needs
a 3D formal solution method of the Stokes-vector 
transfer equation. This is because here the absorption matrix  
turns out to be a full $4\times4$ matrix and all the Stokes parameters
are coupled together. To this end
we have generalized to 3D the Stokes-vector formal solver developed
by Trujillo Bueno (1998), which can be considered as a generalization
to polarization transfer of the short-characteristics method.

We would like to end this paper by saying that over the last 10 years
we have witnessed impressive progress concerning the development
of highly convergent iterative schemes 
and accurate formal solvers for RT applications.
Now it is time to apply them with physical intuition in order
to improve our knowledge of the Sun, its magnetic field
and its polarized spectrum.

\acknowledgements{\footnotesize}
This work has been partially supported by the Spanish DGES through
project PB 95-0028.

\end{document}